\documentstyle[preprint,aps,psfig]{revtex} 
\def\be{\begin{equation}}
\def\ee{\end{equation}}
\def\bea{\begin{eqnarray}}
\def\eea{\end{eqnarray}}

\begin{document}

\renewcommand{\thefootnote}{\fnsymbol{footnote}} 
\def\scf{\setcounter{footnote}}
\hyphenation{}

\title{ON THE HAMILTONIAN FORMULATION OF CLASS B BIANCHI 
COSMOLOGICAL MODELS \thanks{Based in part on a thesis by S. Waller
submitted to the 
University of Texas at Austin in partial fulfilment of the requirements
for the Ph. D. degree.}}

\author{Michael P. Ryan, Jr. and
Sergio M. Waller} 

\address{Instituto de Ciencias Nucleares, 
Universidad Nacional Aut\'onoma de M\'exico \\ 
A. P. 70-543 04510 M\'exico, D.F, MEXICO}
      
\maketitle
\begin{abstract}
The application of the divergence theorem in a non-coordinated basis is shown
to lead to a corrected variational principle for Class B Bianchi cosmological
models. This variational principle is used to construct a Hamiltonian
formulation for diagonal and symmetric vacuum Class B models. [Note: This
is an unpublished paper from 1984 that might be useful to anyone interested
in Class B Bianchi models].
\end{abstract}

\newpage
\section{Introduction}
\bigskip
For the past decade there has been a low-level but continuing interest in
Bianchi-type cosmological models which admit a cosmic time, a gauge in which
$g_{0i}=0$, and whose t = const. surfaces are homogeneous three-spaces
of Bianchi types I-IX \cite{Bian}. 
This interest, for the most part, centers on the
mathematical structure of the Einstein equations for these models instead of on
astrophysical applications, since these models could only represent an
unobservable portion of the actual universe (although a search for observable
consequences of this primitive stage continues). The advantage of the 
Bianchi-type models is that they 
provide a model of general relativity which has true
dynamics, even though the metric components in the invariant basis are fuctions
of time only. These models allow one to study problems in general relativity in
a situation where they are soluble exactly (or at least qualitatively).

One line of study has been the Hamiltonian formulation of the Einstein
equations for these models which was begun with an eye toward approximate
solutions and the construction of a 
model quantum theory for general relativity.
The Hamiltonian formulation for Types I and IX based on the Arnowitt-Deser
Misner\cite{Arn} (ADM) formulation of general relativity is due to 
Misner\cite{Mis}, and was extended to all models of Ellis-MacCallum 
\cite{Ell} Class A by Ryan \cite{Ry1}. In Ref. \cite{Ry1} an extension was made 
to Class B models, but MacCallum and Taub \cite{Mac} pointed out that the naive 
application of the ADM formalism yields a Hamiltonian formulation that gives 
incorrect Einstein equations for these models. This pathology excited some 
interest, and a number of corrected variational principles have been proposed 
for these models \cite{Sne,Gow,Cha,Jan}

In this paper we plan to make a detailed review of the problem of the
Hamiltonian formulation of Class B models from a slightly different point
of view. All of the variational principles in Refs. [7-10] have what Jantzen
\cite{Jan} calls ``non-potential" terms in their variational principles,
and 
there is at least one other problem that these models have that makes the study 
of this Hamiltonian formulation interesting. We plan to emphasize that the 
non-potential terms depend on the treatment of certain terms in the Einstein 
action, and that no extra correction terms are needed. In fact, with the proper 
treatment of the divergence theorem in a non-coordinated basis, the correction 
becomes obvious. This correction provides a foundation for the idea of one of 
us \cite{Ry2} that the problem of Class B models is due to problems of 
variational principles in a non-coordinated basis, although not in the form 
originally proposed.

A model problem for Class B models that suffers from all the same problems
(although in a slightly different form) is that of conformal metrics in general
relativity. Consider the metric
$$
ds^{2} = e^{2\lambda(x,t)}(\eta_{\mu\nu}dx^{\mu}dx^{\nu}). \eqno(1.1)
$$
The Einstein tensor is
$$
G_{\mu\nu}=2\partial_{\mu}\partial_{\nu}\lambda -
\partial_{\mu}\lambda\partial_{\nu}\lambda -
2\eta_{\mu\nu}(\eta^{\alpha\beta}\partial_{\alpha}\partial_{\beta}\lambda +
{{1}\over {4}} 
\eta^{\alpha\beta}\partial_{\alpha}\lambda\partial_{\beta}\lambda). 
\eqno(1.2)
$$
The Einstein action in vacuum becomes
$$
I={1\over {16\pi}} \intop \sqrt{-g} R \, d^{4} x =
{1 \over {8\pi}} \intop e^{2\lambda} [6\eta^{\alpha\beta} \partial_{\alpha}
\partial_{\beta}\lambda +
4\eta^{\alpha\beta}\partial_{\alpha}\lambda\partial_{\beta}\lambda]
d^{4}x. \eqno (1.3)
$$
Here it is easy to see one of the pathologies we will see in Class B models:
{\it Reduced actions may lack equations}. Varying (1.3) with respect to
$\lambda$ gives only one equation, and setting $G_{\mu\nu}$ from (1.2) equal to
zero gives nine more, and only in cases where these nine are automatically
zero or redundant (as in $k=0$ FRW models) will (1.3) yield the correct
Einstein equations.

The second difficulty that we will encounter in Class B models can be
illustrated by rewriting the action (1.3) in an orthonormal basis,
${\bf e}_{\mu} = e^{-\lambda}(\partial/\partial x^{\mu}),\, \omega^{\mu} =
e^{\lambda}dx^{\mu}$. We will write the action of ${\bf e}_{\mu}$ on a function
$A$ as ${\bf e}_{\mu}A \equiv A,_{\mu} = 
e^{-\lambda}\partial_{\mu}A$. The volume
element $d^{4}x = e^{-4\lambda}\omega^{0}\wedge\omega^{1}\wedge\omega^{2}
\wedge\omega^{3}$, so the action becomes
$$
I = {1\over {8\pi}} \intop [6\eta^{\alpha\beta}\lambda_{,\alpha,\beta} +
10\eta^{\alpha\beta}\lambda_{,\alpha}\lambda_{,\beta}]
\omega^{0}\wedge\omega^{1}\wedge\omega^{2}\wedge\omega^{3}. \eqno (1.4)
$$
If one treats the derivatives ${\bf e}_{\mu}$ like 
ordinary derivatives and varies
(1.4), one finds as the vacuum equations
$$
\eta^{\alpha\beta}\lambda_{,\alpha,\beta} = 0 =
\eta^{\alpha\beta}\partial_{\alpha}\partial_{\beta}\lambda -
\eta^{\alpha\beta}\partial_{\alpha}\lambda\partial_{\beta}\lambda .
\eqno (1.5)$$
Varying (1.3) one finds
$$
6\eta^{\alpha\beta}\partial_{\alpha}\partial_{\beta}\lambda +
4\eta^{\alpha\beta}\partial_{\alpha}\lambda\partial_{\beta}\lambda .
$$

The obvious problem is treating ${\bf e}_{\mu}$ derivatives as though they were
partial derivatives in integration by parts. This is the second problem
encountered in Class B models: {\it Integration by parts in non-coordinated
bases can lead to incorrect equations}. In Sec. II we will show that the
answer to this problem lies in the proper use of the divergence theorem in
a non-coordinated basis based on the model of Spivak \cite{Spi} or
Lovelock and Rund \cite{Lov}.

The main thrust of this paper is to show how the application of the correct
divergence thoerem can help in the understanding of the non-potential terms
in the Hamiltonian formulation of Class B models. In fact, the only
non-potential terms are those that arise from the derivatives of the
connection coefficients in the Ricci scalar in the Einstein variational
principle. As a practical excercise we apply the results to diagonal and
symmetric \cite{Go} Class B metrics. While we feel that our presentation
is more didactic, some of the results for diagonal and symmetric models can 
be found in the exhaustive review of Jantzen \cite{Jan}. Note that Jantzen
manages to achieve a Hamiltonian formulation without non-potential terms 
valid for certain metric variables, at the cost (admmittedly slight) of
introducing $g_{0i}$ terms which must be found by integrating 
supplementary equations.

The plan of the paper is as follows. In Sec. II we discuss the divergence
theorem in a non-coordinated basis and its application to the Hamiltonian
formulation of Class B models. In Sec. III we consider vacuum Class B models
with diagonal space metrics, showing which of them allow solutions. Certain
models do not, and in Sec. IV we show that all models with symmetric metrics
do allow vacuum solutions.
\bigskip

\section{The Divergence Theorem in a Non-Coordinated Basis and the Variational
Principle for Class B Models.}
\bigskip

As mentioned in the introduction the corrected variational principles that have
been proposed for Class B models [7-10] have non-potential terms
(with the exception of the final Hamiltonian of Jantzen \cite{Jan}). The first
calculation of correction terms was due to Sneddon \cite{Sne}, who called, the
non-potential terms ``surface terms". Our approach is slightly different, and,
we feel, somewhat more didactic.

The basic cause of the problem for Class B models is that the divergence 
theorem in a non-coordinated basis \cite{Spi,Lov} does not have the usual
form. If we take, for example, the formulation of Lovelock and Rund \cite{Lov},
we find that the divergence theorem, a special case of Stoke's theorem, takes
the following form in an n-dimensional space. If we have a vector $A = A^{i}
{\bf e}_{i}$, the divergence theorem is given in 
terms of an $(n-1)-$ form $\pi_{j}$
defined by
$$
\pi_{j} \equiv (-1)^{j+1} \omega^{1}\wedge \cdots \wedge \omega^{j-1} \wedge
\omega^{j+1} \wedge \cdots \wedge \omega^{n}
\eqno (2.1)
$$
Stokes theorem for the form $\sigma = \pi_j A^j$ is
$$
\int\limits_{G} d\sigma = \int\limits_{\partial G} \sigma . \eqno (2.2)
$$
If we calculate $d\sigma$ we find 
$$
d\sigma = A^{i}_{,i} \omega^{1}\wedge
\cdots \wedge\omega^{n} + A^{j}(-1)^{j+1}d \omega^{1}\wedge\cdots\wedge
\omega^{j-1}\wedge\omega^{j+1}\wedge\cdots\wedge\omega^{n} 
$$
$$
+ \cdots + A^{j}(-1)^{j+1}\omega^{1}\wedge\cdots\wedge\omega^{j+1}\wedge\cdots
\wedge d\omega^{n},
$$
where $A^{i}_{,j} \equiv {\bf e}_{j}A^{i}$. 
Using the fact that in the terms containing
$A^{j}$ and $d\omega^{k}$ only the 
terms ${{1}\over {2}} C^{k}_{kj}\, \omega^{k}\wedge
\omega^{j}$ give a non-zero contribution, we find that
$$
d\sigma = (A^{i}_{,i} + A^{i}C^{l}_{jl}) \omega^{1}\wedge\cdots\wedge
\omega^{n}. \eqno (2.3)
$$
This means that
$$
\int\limits_{G} A^{i}_{,i} \omega^{1}\wedge\cdots\wedge\omega^{n} =
\int\limits_{G} -A^{i}C^{l}_{jl} \omega^{1}\wedge\cdots\wedge\omega^{n} +
\int\limits_{\partial G} \sigma \eqno (2.4)
$$
It is a moot point whether first term on 
the right-hand-side of (2.4) should be called a
``surface term" or not. Here we have chosen to say that only the integral
over $\partial G$ is to be regarded as a surface term.

We would now like to apply the above form of the divergence theorem to Class
B Bianchi models. If we write the Einstein action for Bianchi models in the
basis $(d\tau , \omega^{1}, \omega^{2}, \omega^{3})$, where $\tau$ is a time
we choose and $\{\omega^{i}\}$ is the invariant basis, we find that
$$
I={1\over {16\pi}} \intop {^{4}R}\sqrt{-^{4}g} \, d\tau\wedge\omega^{1}
\wedge\omega^{2}\wedge\omega^{3}, \eqno (2.5)
$$
where ${}_{}^{4}g$ is the determinant of the metric in the above basis. The
usual ADM reduction gives
$$
I = {1 \over {16\pi}} \intop [\pi^{ij} \dot g_{ij} - N{\cal{H}}_{\perp} -
N_{i}{\cal{H}}^{i}] d\tau\wedge\omega^{1}\wedge\omega^{2}\wedge\omega^{3},
\eqno (2.6)
$$
where the $\pi^{ij}$ are treated as an 
independent momenta, $g_{ij}$ is the metric
on $\tau =$ const. surfaces in the basis $\{\omega^{i}\} , N_{i} =
g_{0i}$, and $N = 1/(-g^{00})^{1/2}$. The Hamiltonian and space constraints
are
$$
{\cal{H}}_{\perp} \equiv -\sqrt{g} \{R + g^{-1}[{{1}\over {2}}(\pi^{k}k)^{2} 
- \pi^{ij} \pi_{ij}]\},
$$
$$
{\cal {H}}^{i} \equiv -2(\pi^{ij}_{,j} + \Gamma^{i}_{jk} \pi^{jk} +
\Gamma^{ij}_{kj} \pi^{ik} -
\Gamma^{k}_{kj} \pi^{ij}), \eqno (2.8)
$$
where, as usual, three-dimensional indices are raised and lowered with
$g_{ij}$ and $g^{ij} \equiv [g_{ij}]^{-1}$, $g \equiv 
{\rm det} [g_{ij}]$, comma $i$
means operation with ${\bf e}_{i}$, the 
invariant vector dual to $\omega^{i}$, and
$R$ is the three-dimensional Ricci scalar \cite{Ry3},
$$
g^{rs} (\Gamma^{l}_{rs})_{,l} - g^{rs} (\Gamma^{l}_{rl})_{,s} +
g^{rs} \Gamma^{t}_{rs} \Gamma^{l}_{tl} - g^{rs} \Gamma^{t}_{rl} \Gamma^{l}_{ts}
- g^{rs} C^{t}_{ls} \Gamma^{l}_{rt}. \eqno (2.9)
$$
We would like to assume that the Bianchi models are spatially homogeneous,
that is that all quantities appearing in the action have no space derivatives.
Because of the changed divergence theorem this is not possible, so we will
attempt a ``maximal homogenization" compatible with Class B models. Spatial
derivatives appear in three places: i)$\pi^{ij}_{,j}$, ii) 
$\Gamma^{i}_{jk,l}$,
and iii) the spatial derivatives of the $g_{ij}$ that appear in the
expresion for $\Gamma^{i}_{jk,}$
$$
\Gamma^{i}_{jk} = {{1}\over {2}}g^{il}(g_{lj,k} + g_{lk,j} - g_{jk,l})
$$
$$
+ {{1}\over {2}} (C^{i}_{jk} - g_{jp}g^{in}C^p_{nk} - g_{pk}g^{in}C^p_{nj}). 
\eqno (2.10)
$$

We will show that is possible to ignore the $g_{ij,k}$ 
in the $\Gamma^{i}_{jk,}$
and use in the final action only $\tilde\Gamma^{i}_{jk} = 
{{1}\over {2}}(C^{i}_{jk} -
g_{jp} g^{in}C^p_{nk} - g_{pk}g^{in}C^p_{nj})$. 
From the form of the action (2.7)
and of ${\cal{H}}^{i}$, it is obvious that the term with $\pi^{ij}_{,j}$ will
cause no problem in any gauge in which $N_{i} = 0$, and we will always work
in such a gauge. All of this means that only the terms in $R$ that depend on
derivatives of the $\Gamma^{i}_{jk}$ will cause problems. It has been shown
\cite{Ry2} that the problem with $R$ without derivative terms is that
$\delta (\sqrt{g}R) / \delta g_{ij} \neq - \sqrt{g}R^{ij} + {{1}\over {2}}
\sqrt{g} g^{ij}R$. We will show that if $R$ is given by (2.9) and we use
the divergence theorem (2.5), that indeed $\delta (\sqrt{g}R)/\delta g_{ij} =
- \sqrt{g}R^{ij} + {{1}\over {2}} 
\sqrt{g} g^{ij}R$. We will use $\tilde{R}$ to denote $R$
written with $\tilde{\Gamma}^{i}_{jk}$ in place of $\Gamma^{i}_{jk}$. If we
set the derivative terms in $R$ equal to zero we find that
$$
R(\Gamma_{,} = 0) = - {{1}\over {2}}C^{t}_{si} C^{s}_{tj} g^{ij} -
{{1}\over {2}}C^{a}_{si} C^{b}_{tj} g^{st} g_{ab} g^{ij} 
$$
$$
+ {{1}\over {4}}g^{st} g^{pr} C^{j}_{sp} C^{b}_{tr} g_{bj} +
C^{s}_{sr} C^{k}_{tk} g^{rt}. \eqno (2.11)
$$
When we vary $R$ we will homogenize after variation, so after variation all
derivative terms will be put equal to zero. If in (2.9) we replace
$\Gamma^{i}_{jk}$ by $\tilde\Gamma^{i}_{jk}$, we can see that the term
$g^{rs}(\tilde\Gamma^{l}_{rl})_{,s}$ will not 
contribute to the variation since
$\tilde\Gamma^{l}_{rl} = C^{l}_{rl}$ and is independent of $g_{ij}$. Variation now
gives
$$
{\delta (\sqrt{g}R) \over \delta g_{ij}} =
{{1}\over {2}}\sqrt{g} g^{ij}R +
\sqrt{g} \left({\delta R[\Gamma_{,} =0] \over \delta g_{ij}}\right) -
\sqrt{g} g^{rs}C^{n}_{ln} \left({\delta\tilde\Gamma^{l}_{rs} \over
\delta g_{ij}}\right). \eqno (2.12)
$$
Using (2.11) and the definition of $\tilde\Gamma^{i}_{jk,}$ it is not
difficult to show that
$$
{\delta (\sqrt{g} R) \over \delta g_{ij}} =
{{1}\over {2}}\sqrt{g} g^{ij}R - \sqrt{g} R^{ij}, \eqno (2.13)
$$
and since in the spatially homogeneous case $\tilde{R} = R$ we see that our
assertion that variation of $R$ using the proper divergence theorem gives the
correct Einstein equations.

In the following two sections we will apply the above formalism to the
simplest classes of vacuum models and construct a Hamiltonian formalism. In
our case we can follow the outline given in Ryan and Shepley \cite{Ry3},
where we use the parametrization
$$
g_{ij} = e^{2\Omega}(e^{2\beta})_{ij}, \eqno (2.14)
$$
where $\Omega = \Omega (t)$ and the matrix $\beta_{ij}$ is
$$
\beta = e^{-\psi \kappa^{3}} e^{-\theta \kappa^{1}} e^{-\phi \kappa^3}
\beta_{d} e^{\phi \kappa^{3}}
e^{\theta \kappa^{1}} e^{\psi \kappa^{3}} \eqno (2.15)
$$
where
$$
\kappa^3 = \left[ \matrix{0&1&0\cr
                         -1&0&0\cr
                         0&0&0\cr}\right ], \qquad \kappa^1 
                         = \left [\matrix{0&0&0\cr
                                          0&0&1\cr
                                          0&-1&0\cr}\right ], \eqno (2.16)
$$
and
$$
\beta_{d} = {\rm diag} (\beta_{+} +\sqrt{3}\beta_{-}, 
\beta_{+} - \sqrt{3}\beta_{-}, -2\beta_{+}). \eqno (2.17)
$$
we also define a matrix $p_{ij}$ as
$$
p_{ij} \equiv 2\pi(e^{\beta}_{is} \pi^{s}_{t} e^{-\beta}_{tj})
-{{2\pi} \over 3} \delta_{ij} \pi^{k}_{k}, \eqno (2.18)
$$
where $p_{ij}$ is a function of $(p_{\pm}, p_{\phi}, p_{\psi}, p_{\theta},
\beta_{\pm}, \theta , \phi , \psi )$ such that
$$
I = \intop [p_{+} d{\beta_{+}} +p_{-}d{\beta_{-}} +p_{\psi}d{\psi}
+p_{\phi}d{\phi} + p_{\theta}d{\theta} - Hd\Omega]. \eqno (2.19)
$$
The Hamitonian $H$ is a function of $(p_{\pm} , p_{\phi}, p_{\psi}, p_{\theta},
\beta_{\pm} , \theta , \phi , \psi)$ equal to $2\pi (\pi^{k}_{k})$ and is
obtained by solving ${\cal{H}}_{\perp} = 0$. In those cases where the 
${\cal{H}}^{i}$ are not automatically zero, we add the additional constraints
${\cal{H}}^{i} = 0$. It is not difficult to show that for all Class B models
${\cal{H}}^{i} \neq 0$ and
$$
H^{2} =  6p_{ij}p_{ij} - 24{\pi}^{2} e^{-6\Omega}\tilde{R}. \eqno (2.20)
$$
The ``potential" term now has two parts, $R(\Gamma, = 0)$ and
$g^{rs} (\tilde{\Gamma}^{l}_{rs,l})$ and the Einstein equations are
Hamilton's equations in the following form:
$$
\dot{q} = \delta H/\delta p, \qquad \dot{p} = -\delta H/\delta q, \eqno (2.21)
$$
where $ \cdot \equiv d/d\Omega$ and 
the $\delta$ derivatives include integration
by parts by means of the correct divergence theorem. An important point is
that the term $g^{rs}(\Gamma^{l}_{rs,l})$ can never be converted into a
pure potential term, because the part of the variation of the form
$\delta g^{rs}(\Gamma^{l}_{rs,l})$ is set equal to zero, while
$g^{rs}(\delta\Gamma^{l}_{rs,l})$ gives a non zero term. There does not
exist any pure function of $g_{ij}$ which has this property. Note also that
a good test of the correctness of (2.21) will be $({\cal H}^{i})^{\cdot} =
0$        
as an identity.

In the following two sections we shall study the diagonal ($\beta$ diagonal)
and symmetric cases ($\beta$ with one off-diagonal term) to show effect of
the term $g^{rs}(\Gamma^{l}_{rs,l})$ on Hamilton's equations.
\bigskip
\section{Diagonal Vacuum Class B Models}
It is worthwhile to consider the vacuum diagonal Class B models as a 
cautionary tale, because they are a zoo of pathology. In principle it is
easy to write the corrected Hamiltonian for the Class B models. We have
$p_{ij} = p_{+}\alpha_{1} + p_{-}\alpha_{2}$, where
$$
\alpha_1 = \left [ \matrix {1&0&0\cr
                            0&1&0\cr
                            0&0&-2\cr}\right ] \qquad
\alpha_2 = \left [ \matrix {\sqrt{3}&0&0\cr
                            0&-\sqrt{3}&0\cr
                            0&0&0\cr}\right ]. \eqno (3.1)
$$
The squared Hamiltonian becomes
$$
H^{2} = p_{+}^{2}+p_{-}^{2} +
12\pi^{2} e^{-4\Omega} V(\beta_{\pm})
- 24\pi^{2} e^{-6\Omega} g^{rs}(\tilde{\Gamma}^{l}_{rs,l}). \eqno (3.2)
$$
We use the form of the Ellis-MacCallum \cite{Ell} scheme for writing the
constant $C^{i}_{jk}$ given in Ryan and Shepley \cite{Ry3}. That is,
$$
C^{i}_{jk} = \varepsilon_{jks}m^{si} +
\delta^{i}_{k} a_{j} -
\delta^{i}_{j} a_{k}.
$$
For all Class B Models $a_{i} = (1/2)C^l_{il} \propto \delta^{3}_{i}$, so
only the term $g^{rs} (\tilde\Gamma^{3}_{rs})_{,3}$ contributes to $H^{2}$.
Since $m^{ij}a_{j} = 0$ by the Jacobi identity, $m^{i3} = 0$ for all $i$.
If we define the {\it spin coefficients} of the matrix $m^{AB}, A, B =1, 2$
as $\alpha = {{1}\over {2}}(m^{11}+m^{22})$, 
$\lambda = -{{1}\over {2}}(m^{11} - m^{22})$, $\gamma =
m^{12}$, the Hamiltonian (3.2) reduces to
$$
H^{2} = p_{+}^{2} +p_{-}^{2} +
12\pi^{2} e^{-4\Omega}e^{4\beta_{+}}[2(\alpha^{2} + \lambda^{2})
\cosh \ (4\sqrt{3} \beta_{-}) 
$$
$$
+ 4\lambda\alpha \sinh \ (4\sqrt{3} \beta_{-})
-2\alpha^{2} + 2\lambda^{2} + 4\gamma^{2} + 12a_{3}^{2}]
-24\pi^{2} e^{-4\Omega}e^{4\beta_{+}} [-12a_{3}(\beta_{+})_{,3}
-4\sqrt{3}\gamma(\beta_{-})_{,3}], \eqno (3.4)
$$
where we have replaced $\tilde\Gamma^{l}_{rs}$ by its expression in terms of
$g_{ij}$ in the parametrization (2.14) and taken the necessary derivatives.
Notice that the first of the derivative terms 
depends only on $\beta_{+}$, and this
allows one to integrate by parts.  That is, using the correct divergence 
theorem, variation of the part of $H$
containing $a_{3}$ gives
$$
{\delta H\over \delta\beta_{+}} (\alpha =\lambda =\gamma =0) =
{1\over H} \{576\pi^{2} e^{-4\Omega}e^{4\beta_{+}} a_{3}^{2} -
576\pi^{2} e^{-4\Omega} e^{4\beta_{+}}a_3^2\} = 0, \eqno (3.5)
$$
so a reduced Hamiltonian  for all diagonal Class B models which gives the
equations of motion for $\beta_{\pm}$ (i.e. $\dot{H} \neq \partial H/
\partial \Omega)$ is
given by
$$
H^{2} = p_{+}^{2} +p_{-}^{2} + 12\pi^{2} e^{-4\Omega} e^{4\beta_{+}}
[2(\alpha^{2}+\lambda^{2}) \cosh (4\sqrt{3}\beta_{-})
+ 4\lambda\alpha \sinh (4\sqrt{3}\beta_{-}) -
$$
$$
-2\alpha^{2} +2\lambda^{2} +4\gamma^{2}] + 
96\sqrt{3} \pi^{2} e^{-4\Omega} e^{4\beta_{+}} \gamma(\beta_{-})_{,3}.
\eqno (3.6)$$
This reduction was also noticed by Jantzen \cite{Jan}. No further integration
by parts is possible, because variation of the 
remaining derivative term with respect to
$\beta_{+}$ must give zero and with respect to $\beta_{-}$ must give
$- 192 \sqrt{3}\pi^{2} e^{-4\Omega} e^{4\beta_{+}} \gamma$.

To study this Hamiltonian we give $\alpha$,
$\lambda$, $\gamma$, and $a_3$ in Table I.
It would {\it seem} at this point that Types V and VI give simple Hamiltonians
because $\gamma = 0$. However, it is instructive to apply the test suggested in
the previous section, that is, $({\cal H}^{i})^{\cdot} = 0$. For all the
Class
B diagonal models ${\cal H}^{1}$ and ${\cal H}^{2}$ are identically zero
and ${\cal H}^{3} = 0$ reduces to
$$
6a_{3}p_{+} +2\sqrt{3}\gamma p_{-} = 0 \eqno (3.7)
$$
\vfil\eject

\centerline {Table I}
\vskip 10 pt
\halign{\qquad\qquad\hfil 
#&\qquad\qquad\hfil#\hfil&\qquad\qquad\hfil#\hfil&\qquad\qquad\hfil#\hfil&
\qquad\qquad#\hfil\cr
Bianchi Type&$\alpha$&$\lambda$&$\gamma$&$a_3$\cr
III&$0$&$0$&${{1}\over {2}}$&$-{{1}\over {2}}$\cr
IV&${{1}\over {2}}$&$-{{1}\over {2}}$&$0$&$-1$\cr
V&$0$&$0$&$0$&$-1$\cr
VI$_{h \neq -1}$&$0$&$0$&${{1}\over {2}}(h-1)$&$-{{1}\over {2}}(h + 1)$\cr
VII$_{h \neq 0}$&$-1$&$0$&${{h}\over {2}}$&$-{{h}\over {2}}$\cr}
\vskip 10 pt

It is easy to see that Type V models with $\alpha = \gamma = 0$ are consistent.
The reduced Hamiltonian is
$$
H^{2} = p_{+}^{2} +p_{-}^{2}, \eqno (3.8)
$$
and $p_{+} = 0$ is the space constraint. It is even possible to construct
a Hamiltonian that obeys $\dot{H} = \partial H/\partial\Omega$, that is,
$p_{+} = 0$ implies $\beta_{+} = \beta_{+}^{0}$ and a final Hamiltonian is
$$
H^{2} = p_{-}^{2} +12\pi^{2} e^{-4\Omega} e^{4\beta_{+}^{0}}. \eqno (3.9)
$$
This Hamiltonian gives as a general solution for vacuum diagonal Type V models
$$
\beta_{+} = \beta_{+}^{0}
$$
$$
p_{-} = {\rm const.}
$$
$$
\beta_{-} = \beta_{-}^{0} +p_{-}\Omega
$$
$$
H = \{p_{-}^{2} +12\pi^{2} e^{-4\Omega} e^{4\beta_{+}^{0}}\}^{1/2} =
{{1}\over {12\pi}} e^{3\Omega} \left({d\Omega \over dt}\right)^{-1} 
\eqno (3.10)
$$
Let us now examine the Type IV case. If we calculate $6a_{3}\dot{p}_{+}$
$(\gamma = 0)$ we find
$6a_{3}\dot{p}_{+} = 6a_{3} \left[48\pi^{2} e^{-4\Omega} e^{4\beta_{+}}
\{\cosh (4\sqrt{3} \beta_{-}) \ - \sinh (4\sqrt{3} \beta_{-})\}\right] 
\not= 0$ !

This equation is one of the reasons we have qualified the diagonal vacuum
Class B models as a zoo of pathology. A check of $({\cal H}^{i})^{\cdot} =
0$
shows that models of types III, V, and VI$_{h \neq -1}$ are consistent, that
is those for which $\alpha = \lambda = 0$. The two remaining types, IV and
VII$_{h \neq 0}$, are examples of another problem in reduced variational
principles. We have written our variational principle for diagonal metrics
by inserting a diagonal metric directly into the Einstein action. The problem
here is that if the $R_{ij}$, $(i \neq j)$ 
are not identically zero, we will lose the
equations $R_{ij} = 0$ $(i \neq j)$. A rapid calculation shows that $R_{ij}$ is
diagonal for a diagonal metric for types III, V, and VI$_{h \neq -1}$, while
for types IV and VII$_{h \neq 0}$,  $R_{12} \neq 0$. 
This seems to indicate that
symmetric metrics $(g_{12} \neq 0)$ will always provide consistent vacuum
equations if such metrics always have $R_{13} = R_{23} = 0$ identically. In
the following section we show that this is indeed the case, and present a
consistent Hamiltonian formulation for all symmetric vacuum Class B models.
\bigskip
\section{The Symmetric Case}
\bigskip
In the previous section we have shown that the corrected Hamiltonian as given
by (2.20) does not always yields equations of motion which are consistent
with the space constraint equation when one assumes that the metric of the
homogeneous hypersurface is diagonal. This is certainly the case with diagonal
metrics that admit groups of isometries 
of Bianchi types IV and VII$_{h \neq 0}$.
We found out that a common feature appeared in those cases where the 
Hamiltonian equations of motion were not 
consistent with the space constraint. Namely, that
some of the non-diagonal components of the Ricci tensor do not vanish as one
might expect from the diagonality of the space metric.

In this Section we shall consider symmetric metrics (one nonzero off-diagonal
element in the metric) and show that all the components of the Ricci tensor
for which the corresponding metric component is zero vanish in this case.
Therefore the ``pathology" found with diagonal metrics is not present in the
symmetric case, which tells us that the corrected Hamiltonian for symmetric
metrics must provide the precise equations of motion that are consistent with
the space constraint equation. This is formally shown by writing explicity
the
corrected Hamiltonian for symmetric metrics and the corresponding Hamilton
equations. Then these equations are used along with the space constraint
equation
to show that the latter is fulfiled at all times, and therefore that the
equations of motion are consistent with the space constraint equation.

As we are assuming a space symmetric metric, we set $g_{13} = g^{13} = g_{23} =
g^{23} = 0$ and then we have only to compute the $R_{13}$ and $R_{23}$
components of the Ricci tensor in the order to find out whether they vanish or
not. The Ricci tensor is given by 
$$
R_{ij} = {{1}\over {2}} a_{i}a_{j} +
2g_{ij} g^{kt}a_{k}a_{t} +
g_{ik} g^{sb} \varepsilon_{sjl} m^{lk} a_{b}
$$
$$
+g_{jk} g^{sb} a_{b} \varepsilon_{sil} m^{lk} a_{b} -
1/2 \varepsilon_{sjt} \varepsilon_{kin} m^{tk} m^{ns}
$$
$$
- {{1}\over {2}} \varepsilon_{sjt} 
\varepsilon_{kin} g^{sk} g_{ab} m^{tb} m^{na},
\eqno (4.1)
$$
with $a_{i}$, $m^{ij}$, $g_{ij}$ and $\varepsilon_{ijk}$ as defined previously.
Then immediately follows that the $R_{13}$ component is
$$
R_{13} = g_{3k} g^{33} m^{2k} a_3^{2} -
{{1}\over {2}} \varepsilon_{13t} \varepsilon_{k1n} m^{tk} m^{n1} -
{{1}\over {2}} \varepsilon_{23t}\varepsilon_{k1n} m^{tk} m^{n2}- 
$$
$$
- {{1}\over {2}} \varepsilon_{13t}\varepsilon_{k1n}m^{tb}m^{na} -
{{1}\over {2}} \varepsilon_{23t}\varepsilon_{k1n}
g^{2k}g_{ab}m^{tb}m^{na}, \eqno (4.2)
$$
where we have used the fact that $a_{1} = a_{2} = 0$ (see Ref. [15]),
$g_{13} = 0$ and the symmetry and skewsymmetry of $g^{ij}$ and $\varepsilon
_{ijk}$ respectively. Expression (4.2) can be reduced to
$$
R_{13} = 1/2 \varepsilon_{k1n} g^{1k} g_{ab} m^{2b} m^{na},
$$
and it is straightforward to show that this $R_{13} = 0$.

Now we compute the $R_{23}$ component of the Ricci Tensor. By the same 
arguments (viz. $a_{1} = a_{2} = 0$, $g_{23} = 0$, 
$g_{ij}$ and $\varepsilon_{ijk}
= -\varepsilon_{jik}$) from (4.1) we obtain
$$
R_{23} = g_{3k} g^{s3} \varepsilon _{s2l} m^{lk} a^{2}_{3}
- \varepsilon_{13t} \varepsilon_{k2n} m^{tk} m^{n1}
- \varepsilon_{23t} \varepsilon_{k2n} m^{tk} m^{n2}
$$
$$
- {{1}\over {2}} \varepsilon_{13t}\varepsilon_{k2n} g^{lk} g_{ab} m^{tb} m^{na}
- {{1}\over {2}} \varepsilon_{23t} 
\varepsilon_{k2n} g^{2k} g_{ab} m^{tb} m^{na}. 
\eqno (4.4)
$$
Imposing $m^{13} = m^{31} = m^{23} = m^{32} = 0$ and $g_{13} = g_{23} = 0$,
(4.4) becomes
$$
R_{23} = \varepsilon_{k2n} m^{2k} m^{n1} - \varepsilon_{k2n} m^{1k} m^{n2},
\eqno (4.5)$$
or simply $R_{23} = 0$.

We should point out that the above results, namely $R_{13} = 0$ and $R_{23}
= 0$, hold for all symmetric metrics which admit groups of isometries of Class
B.

The form of $g_{ij}$ for the symmetric case is (2.14) and (2.15) with
$\psi = \theta = 0$. The corrected Hamiltonian for this case is
$$
H^{2} = p_{+}^{2} + p_{-}^{2} + {3p_{\phi}^{2} \over \ \sinh (2\sqrt{3}
\beta_{-})} - 24\pi^{2} e^{-6\Omega}\tilde{R}, \eqno (4.6)
$$
where $\tilde R$, as before, is given by
$$
\tilde{R} = R + g^{rs} (\tilde{\Gamma}^{l}_{rs})_{,l}, \eqno (4.7)
$$
with the Ricci (curvature) scalar defined by (2.11) and the homogenized
connections as defined in Sec. II.

As we did in the diagonal case, one can show that the only contribution from
$g^{rs}(\tilde{\Gamma}^{l}_{rs})_{,l}$ comes from the term $g^{rs}
(\tilde{\Gamma}^{3}_{rs})_{,3}$, that is,
$$
g^{rs}{(\tilde{\Gamma}^{l}_{rs})}_{,l} =
- a_{3} [g^{11}(g^{33}g^{11})_{,3} + g^{22}{(g^{33}g_{22})}_{,3}
$$
$$
+ 2g^{12}(g^{33}g_{12})_{,3}] - g^{11} [g^{33}g_{1p}m^{2p}]_{,3}
$$
$$
+ g^{22}[g^{33}g_{2p}m^{1p}] + g^{12}[g^{33}g_{1p}m^{1p}]_{,3}
$$
$$
-g^{12}[g^{33}g_{2p}m^{2p}]_{,3}. \eqno (4.8)
$$
Of course we have used $g_{13}=g_{23}=g^{23}=0, a_{1}=a_{2}=0$ and the
symmetry of $m^{ij}$. 

Now we proceed to obtain the explicit form of (4.8) in terms ot the 
parametrization (2.14 - 2.15). However, in order to have a compact expression
for the final result we introduce the spin coefficients $\alpha,\lambda,\gamma$
of the rotated matrix $e^{\phi \kappa^3}\tilde{m} e^{-\phi \kappa^3}$, 
where $\tilde{m}$ is
given by
$$
\tilde m = \left [ \matrix{m^{11}&m^{12}\cr
                           m^{21}&m^{22}\cr}\right ]
$$
We find
$$
\alpha = {{1}\over {2}} (m^{11} + m^{22}), \eqno (4.9)
$$
$$
\lambda = {{1}\over {2}} \cos 2\phi (m^{11} - m^{22}) + \sin 2\phi m^{12}, 
\eqno (4.10)
$$
and
$$
\gamma = - {{1}\over{2}} \sin 2\phi (m^{11} - m^{22}) + \cos 2\phi m^{12}.
\eqno (4.11)
$$
After a lengthy algebraic computation we end up with the following result
for the expression (4.8),
$$
g^{rs}(\tilde{\Gamma}^{l}_{rs})_{,l} = e^{2\Omega}e^{4\beta+}
[-12a_{3} \beta_{+,3} - 4\sqrt{3} \gamma\beta_{-, 3} - 2\alpha
\phi_{,3}
$$
$$
+ 2\lambda \sinh (4\sqrt{3} \beta_{-})\phi_{,3}
 + 2\alpha \cosh (4\sqrt{3} \beta_{-})\phi_{,3}]. \eqno (4.12)
$$
The scalar curvature $R$ has the same form as in the diagonal case with
$\alpha, \gamma, \lambda$ replaced by (4.9 - 4.11), so the final Hamiltonian
becomes
$$
H^{2} = p_{+}^{2} + p_{-}^{2} + {3p^2_{\phi} 
\over \sinh^{2}(2\sqrt{3}\beta_{-})}
+ 24\pi^{2} e^{-4\Omega}e^{4\beta+} [(\alpha^{2} + \lambda^{2})  
\cosh (4\sqrt{3}\beta_{-}) +
$$
$$
+ 2\alpha\lambda  \sinh(4\sqrt{3}\beta_{-}) - (\alpha^{2} -\lambda^{2})
+ 2\gamma^{2} + 6a_{3}^{2}] -
$$
$$
-24\pi^{2}e^{-4\Omega}e^{4\beta+} [-12a_{3}\beta_{+,3} 
- 4\sqrt{3} \gamma \beta_{-,3} - 2\alpha \phi_{,3}
$$
$$
+ 2\lambda \sinh(4\sqrt{3}\beta_{-})\phi_{,3}
+ 2\alpha \cosh(4\sqrt{3}\beta_{-})\phi_{,3}]. \eqno (4.13)
$$
To compute the space constraint equations 
for this symmetric case we make use of the
expression (2.8) and impose $g_{13}=g^{13}=g_{23}=g^{23}=0$ and $\pi_{13}=
\pi^{13}=\pi_{23}=\pi^{23}=0$.
Equation (2.8) can be broken down into
$$
g^{1t}\pi_{s}^{k}C^{s}_{tk} - C^{j}_{kj} \pi^{1k} = 0, \eqno (4.14)
$$
$$
g^{2t}\pi_{s}^{k}C^{s}_{tk} - C^{j}_{kj}\pi^{2k} = 0, \eqno (4.15)
$$
and
$$
g^{3t}\pi_{s}^{k}C^{s}_{tk} -C^j_{kj}\pi^{3k} = 0 \eqno (4.16)
$$
If one uses the expression in Sec. II for the structure coefficients, then by
virtue of the symmetric form of $g_{ij}$ and $\pi^{ij}$ Equations (4.14) and
(4.15) are satisfied automatically. However equation (4.16) becomes
$$
(\pi^{1}_{1} - \pi^{2}_{2})m^{12} 
- \pi^{2}_{1}m^{12}m^{11}
+ \pi^{1}_{2}m^{22}
$$
$$
+ a_{3}(\pi^{1}_{1} + \pi^{2}_{2} - 2\pi^{3}_{3}) = 0. \eqno (4.17)
$$
The $\pi^{s}_{t}$ can be obtained, as before, from (2.18). In 
the Ryan and Shepley
\cite{Ry3} parametrization the trace free part of $\pi^{s}_{t}$, i.e.
$p_{ij}$, is given by
$$
6p_{ij} = e^{-\phi \kappa^{3}} [\alpha_{1}p_{+} +
\alpha_{2}p_{-} +
\alpha_{3} {3p_{\phi} \over \sinh (2\sqrt{3}\beta_{-})}] e^{\phi \kappa^{3}}, 
\eqno (4.18)
$$
with $\alpha_{1}$ and $\alpha_{2}$ as before and
$$
\alpha_3 = \left [\matrix{0&1&0\cr
                          1&0&0\cr
                          0&0&0\cr}\right ]
$$
It follows from (4.18) that (4.17) can be written as
$$
3a_{3}p_{+} - 3\alpha p_{\phi} -
3\lambda \coth (2\sqrt{3}\beta_{-})p_{\phi} +
\sqrt{3}\gamma p_{-} = 0, \label{4.19}
$$
where we have used the spin coefficients previously defined. Now we can check
that the Hamiltonian (4.13) yields equations of motion which satisfy the
space constraint equation (4.19) at all times. This we can do by taking
the time
derivative of (4.19), that is
$$
3a_{3}\dot{p}_{+} + 2\sqrt{3} \gamma \dot{p}_{-} -
[3\alpha + 3\lambda \coth (2\sqrt{3}\beta_{-})]\dot{p}_{\phi}
$$
$$
+ 6\sqrt{3} \lambda {\rm csch}^{2}
(2\sqrt{3}\beta_{-})p_{\phi}\dot{\beta}_{-}
+ [6\gamma \coth (2\sqrt{3}\beta_{-})p_{\phi} 
- 4\sqrt{3}p_{-}]\dot{\phi}, \eqno (4.20)
$$
where one uses ${d\lambda \over d\phi} = 2\gamma$ and ${d\gamma \over
d\phi} = -2\lambda$. Substituting Hamilton's equations as obtained
from (4.13) for $\dot{p}_{+}, \dot{p}_{-}, \dot{p}_{\phi}, \dot{\beta}_{-}$ and
$\dot{\phi}$ into (4.20), one verifies that it is indeed zero.

\bigskip
\section{Conclusions}
\bigskip
In this paper we have investigated the causes why the variational principles
for Class B Bianchi models breaks down. We have identified as the basic cause
of the problem that the divergence theorem in a non-coordinated bases does not
have the usual form. The corrected form of the divergence theorem has been
derived, and we have shown that applying the usual ADM reduction to the
Einstein action for Bianchi models in combination with this corrected form
of the divergence theorem gives the right Einstein field equations for Class
B models.

The above formalism was used to construct a Hamiltonian formalism for diagonal
($\beta$ diagonal) and symmetric ($\beta$ with one off-diagonal term) vacuum
Class B models.

In the diagonal case we found that the models of Types III, V and
VI$_{h \neq -1}$ are consistent in the sense that the space constraint equation 
is
satisfied at all times. However, Types IV and VII$_{h \neq 0}$ are not. The
problem in the latter two is that although a diagonal metric is inserted
directly into the Einstein action, $R_{12}$ is not identically zero and
therefore the $R_{12} = 0$ equation is lost.

For the symmetric case we showed that $R_{13}$ and $R_{23}$ automatically
vanish as a result of introducing a symmetric metric. Furthermore, all 
vacuum Class B models yield equations of motion which are consistent with
the space constraint equation in the sense previously mentioned.
\bigskip
\section{Acknowledgements}
\bigskip
We wish to thank L. C. Shepley and R. A, Matzner for valuable discussions.

\end{document}